# A UML-based Approach to Design Parallel and Distributed Applications

Yasset Perez-Riverol [a,b], Roberto Vera Álvarez [c]

December 2013

[a] Department of Proteomics, Center for Genetic Engineering and Biotechnology. Ave 31 e/ 158 y 190. Cubanacán, Playa. Ciudad de la Habana. Cuba.

[b] EMBL Outstation, European Bioinformatics Institute, Wellcome Trust Genome Campus, Hinxton, Cambridge, UK.

[c] Departament of Protein Structure and Bioinformatics, International Centre for Genetics Engineering and Biotechnology (ICGEB), Trieste, Italy,

**Abstract**

Parallel and distributed application design is a major area of interest in the domain of high performance scientific and industrial computing. Over the years, various approaches have been proposed to aid parallel program developers to modeling their applications. In this paper it will be used some concepts from agile development methodologies and Unified Modeling Language (UML) to modeling parallel and distributed applications. The UML-based approach of this paper describes through different artifacts and graphs the main flows of events in the development of parallel and high performance applications. Here, we presented three work flows to describe and to model our parallel program, Domain Model, Design and Modeling and Test. All these phases of the development software allow to programmers convert the requirements of the problem in a good and efficient solution.

**Introduction**

According to Moore law [1], the capabilities of computing hardware increase exponentially in time. At the same time, more and more scientific and engineering problems become amenable to numerical and simulation treatment, and ever more complex and efficient algorithms are being developed to make even a better use of the existing computing power. In fact, parallel and distributed computing is becoming an integral part in several major application domains, such as: big data analysis, medicine, genomics and proteomics research, and image processing [2]. With the advent of fast interconnecting networks of workstations and PCs, it is now becoming increasingly possible to develop high-performance parallel applications using the combined computing powers of these networked-resources. Its consent among most of the software developers of parallel applications that the developing of these algorithms is not simple and this complexity is originated by the accumulation of many low level details related with the parallelism of the sequential algorithms and its instrumentation. The development of a parallel applications demands long periods of code, compilation and execution, going by complex processes of detection of errors and tests [3]. Over the years, developers have tried to use the modeling of the applications through components and visual tools to improve the understanding of complex software systems. Most of these investigations have gone in sense of obtaining patterns of design of parallel programs [4].

The emergence of visual modeling language as UML allows representing distributed and parallel applications through of a group of diagrams [5]. The Unified Modeling Language (UML) is a standard language for specifying, visualizing, constructing, and documenting the artifacts of software system, as well as for business modeling and other non-software systems. The UML represents a collection of the best engineering practices that have been proven successfully in the modeling of large and complex systems. The UML uses mostly graphical notations to express the design of software projects. It represents the different phases or stages of the development software (design, analysis, instrumentation and test). The most important works of the use of visual tools in parallel applications modeling are the Project Prophet [6]. They use UML as modeling language to represent and design parallel and distributed architecture and also denote important concepts of process topology and communications modeling [7]. In this manuscript, we used the Prophet Methodology to modeling parallel and distributed applications; also we use other concepts and phases of the Rational Unified Process (RUP) [8] and Extreme Programing (XP) [9].

**Domain Modeling**

According to Rational Unified Process (RUP) a domain model captures the most important types of objects in the context of business [8, 9]. The domain model represents the 'things' that exist or events that transpire in the business environment. The goal of domain modeling is to provide "the big picture" of the interrelationships among business entities in a complex organization. The Domain Model typically shows the major business entities, their functional responsibilities, and the relationships among the entities. It also provides a high-level description of the data that each entity provides. The Domain Model for parallel applications is more close to the crystal clear 360 exploration concept [10]. The idea is a preliminary feasibility study providing a high-level review of the key issues governing the parallel development efforts. The advantages of using Domain Model for modeling distributed and parallel applications are mainly:

- Giving a conceptual framework of the things in the problem space.
- Helping developers to think focus on semantics.
- Providing a glossary of terms noun based
- Provide a skeleton from the system and their functionalities.

The Domain Model (Fig. 1) allows to developers to know which entities and objects of the problems becomes into possible classes and objects in the system instrumentation.

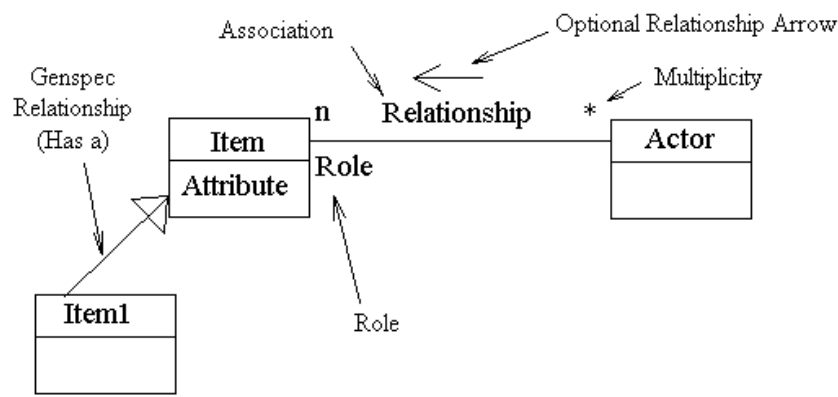

Figure 1: Simple Domain Model Diagram.

The following features enable the developers to express rules for a domain diagram:
- Domain classes: Each domain class denotes a type of object. Attributes: An attribute is the description of a named slot of a specified type in a domain class; each instance of the class separately holds a value.

- Associations: An association is a relationship between two (or more) domain classes that describes links between their object instances. Associations could have roles that describe the multiplicity and participation of a class in the relationship.
- Additional rules: Complex rules that cannot be shown with symbols can be shown with attached notes.

**Software Requirements**

Software requirements describe what the software will do; the software expected environment, the software usage profile, its performance parameters, and its expected quality and effectiveness [11]. They capture the intended behavior of the system. This behavior may be expressed as services, tasks or functions the system is required to perform.

All the software requirements should be represented:
- Correct, unambiguous, complete, and consistent.
- Ranked from importance and/or stability.
- Verifiable, modifiable, and traceable.

In order to represent the software functional requirements, it should write all the operations the program need to obtain from an X input a Y output. It is advisable to distribute tasks in subtasks or sub operations with the purpose of organizing the system and detect the algorithms that can be distributed. For parallel and distributed applications the fundamentals No-Functional requirements are the network architecture and the performance requirements. These conditions have so much to do with the architecture of the network on which the application will be run and the topology used by the software.

**Sequential Algorithm**

After system requirements detection process, the developers have been studied the possible solutions of the problem and also as a product, the sequential flow events that compose de problem solution are known [12]. The programmer has the wished sequential algorithm to parallel and the made operations. A standard form to represent this process is using the natural language describing sequential activities. These descriptions are enough detailed, but they can be difficult to interpret, especially within a complex set of cases of use. Another form to capture those flows is using Activities Diagrams [13]. Those ones describe the flows like a road map of the functional behavior of the system, and display a comprehensible summary of the flow of events within the process, leaving the details of design to other devices. In parallel applications, the activity diagram can be used to describe the sequential algorithm that is desired to parallel as well as the main flows that partake in this.

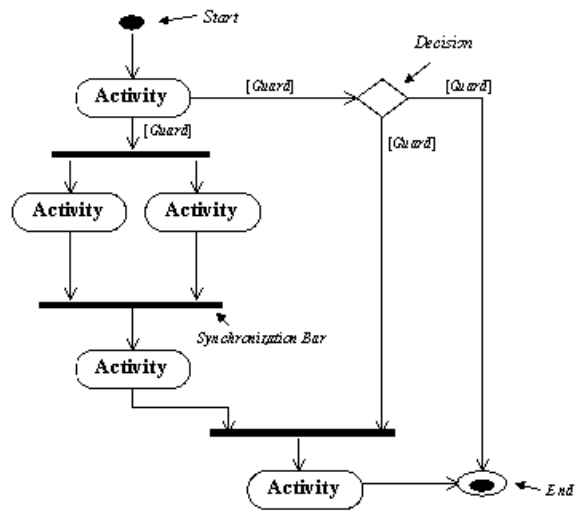

Figure 2: Activity Diagram.

In fact, Activity Diagrams (Fig. 2) can be used for many purposes: diagramming use-case flows; modeling complex business operations or processes; depicting data and information flows; and even computing algorithms. In the related studies the activity diagram allows the developer to identify the parts of the sequential program that can be paralleled.

**Performance Requirements**

The performance requirements are essential in the development of a parallel application. These requirements are very important for choosing which paradigm should be used for modeling the parallel application [14]. In the particular case of the parallel applications, the play is as important as the functional requirements because it justify the existence of a parallel solution on the sequential solution of the problem. They are frequently among the more problematic aspects of parallel program design, but, the good news is that they are those who have easiest formalization in mathematical models. A good performance model, like a good scientific theory, should be able to explain available observations and predict future circumstances, while abstracting unimportant details. We describe the topic of performance requirements by describing some of these mathematical models:

- Process Topology: As a network topology, the process topology is the study of the arrangement or mapping of the elements (links, nodes, etc.) of a network, especially the physical interconnections between nodes. There exist several topologies to communicate the processes: bus, star, ring, mesh, cub, and tree. These topologies are described by many authors; otherwise, several classic algorithms of the sequential programming have been developed in parallel and they also have been described their efficiency on different topologies. All the performance requirements depend of Process Topology; and when the programmer modeled the sequential algorithm to obtain the parallel program is important to study and represent the process topology used.
- Execution Time: The execution time of a parallel program defines itself as the time that elapses from when the first processor starts executing on the problem to when the last processor completes execution. This concept is expressed as the sum of computation, communication.
- Computation time: It depends on processors and memory characteristics.
- Communication time: It is the time that its tasks spend sending and receiving messages.

The decrease of the execution time is the goal of a parallel application. First of all, the program must obtain a result, after; the developers must try to decrease the run time of the sequential program. This time must be sufficiently smaller than the time of the sequential algorithm to justify the implementation of the parallel algorithm. An important aspect of performance requirements analysis is the study of how algorithm performance varies with parameters such as problem size, processor count, and message startup cost. In particular, it's important the evaluation of the scalability of a parallel algorithm, that is, how effectively it can use an increased number of processors.

**Selecting a design pattern**

In order to help the design of parallel algorithms and to represent the stages of the methodical design the developers creates the parallel programming paradigms [4]. A programming paradigm is a class of algorithms that solve different problems but have the same control structure. In the world of parallel computing there are several authors which present a parallel classification [3, 15]. Several programming paradigms are commonly used to develop parallel programs on distributed clusters, for instance:

*Master-Worker (Task-Farming)*: The master-slave paradigm consists of two entities: master and multiple slaves. The master is responsible for decomposing the problem into small task (and distributes tasks among a farm of slave processes). Task-farming may use static load-balancing or dynamic load-balancing. In the first case, the distribution of tasks is all performed at the beginning of the computation. The other way is to use a dynamically load-balanced master/slave paradigm, which can be more suitable when the number of tasks exceeds the number of available processors, or when the number of tasks is unknown at the start of the application, or when the execution times are not predictable, or when we are dealing with unbalanced problems.

An important feature of dynamic load-balancing is the ability of the application to adapt itself to changing conditions of the system, not just the load of the processors, but also a possible reconfiguration of the system resources. This paradigm can be associated to simple problems problem where the sequential algorithm is executed in the slave's processes on different data sets. This paradigm can achieves high computational speedups and interesting degree of scalability.

*Single Program Multiple Data (SPMD)*: The SPMD paradigm is the most commonly used paradigm. Each process executes the same function of code but on different part sets. This involves the splitting of application data among the available processors. This type of parallelism is also referred as geometric parallelism, domain decomposition, or data parallelism. The problems with a regular geometric structure allows to developers distributed the data uniformly across the process. The communication pattern is usually highly structured and extremely predictable. The application based in this model can be very efficient if the data is well distributed by the processes and the system is homogeneous.

*Data Pipelining*: In the pipeline paradigms, the problem is divided into a series of tasks that have to be completed one after the other. This model is a more fine-grained parallelism; which is based on a functional decomposition approach. In fact, this is the basis of sequential programming. The problem is divided into separate functions that must be performed, but in this case, the functions are performed in succession. The communication may be completely asynchronous and the efficiency of this paradigm is directly dependent on the ability to balance the work across the stages of the pipeline.

*Divide and Conquer*: The divide-and-conqueror paradigm is a well-known strategy in the sequential programming. This approach is characterized by dividing a problem into sub-problems that are of the same form as the larger problem. Each of these sub-problems is solved independently and their results are combined to give the final result. The root process first divides the problem into two sub-problems. These two parts are each divided into two parts, and so on until the leaves are reached. In the parallel divide-and-conqueror paradigm, the sub-problems can be solved at the same time in different processes. In this model it can be identified three generic operations: split, compute, and join.

*Hybrid Models*: The limit between the paradigms can sometimes be fuzzy and, in some applications, there could be the need to mix elements of different paradigms. Hybrid methods that include more than one basic paradigm are usually observed in some large-scale parallel applications. These are situations where it makes sense to mix data and task parallelism simultaneously or in different parts of the same program. It's recommended that the developers evaluate the sequential algorithm to find the critical functions of the application. If the limitation of the algorithm is the amount of data that treats, it's recommended to begin by the domain decomposition.

In order to design the application, the developers should review some of these paradigms to be used during the software development process. With the sequential solution previously described and these models the programmer can adapt the problem to the paradigm to obtain the best solution. In this phase of the modeling process the developers should focused in the no-functional requirements. This phase include observations about the process topology, and the performance requirements. The desirable design involves tradeoffs between simplicity, performance and portability.

**Designing with UML**

The Unified Modeling Language (UML) [5] is a non-proprietary specification language for object modeling. UML is a general-purpose modeling language that includes a standardized graphical notation that is good for creating an abstract model of a system, referred to as a UML model. UML is extendable, offering the following mechanisms for customization: profiles and stereotype. The UML language defines nine diagram types that which allow describing different aspects of a system: Use Case Diagram, class diagram, object diagram, state chart diagram, sequence diagram, collaboration diagram, activity diagram, component diagram, and deployment diagram. Each diagram type describes a step of the development of software system. In order to represent and design parallel applications we used a small subset of these artifacts.

The first UML diagram that developers should use to design and describe the parallel solution is the process topology diagram [7, 14, 16]. A process topology may be defined as a group of processes that have a predefined regular interconnection topology such as farm, ring, 2D mesh or tree. The description of a process topology is a machine-independent and depends only on the parallel application that is executed on the certain topology [7]. UML collaboration diagram (interaction diagrams) illustrates the relationship and interaction between software objects. In order to model the process topology, each process represents an object in the collaboration diagram (Fig. 3). In a collaboration diagram, a communication, a message is shown as an arrow attached to a relationship pointing from the sender toward the receiver. In the case of process topology diagrams in this paper was specified with a label the action description and the data that the processes exchange. It is a good practice that the applications that use load-balance represent all the used topologies.

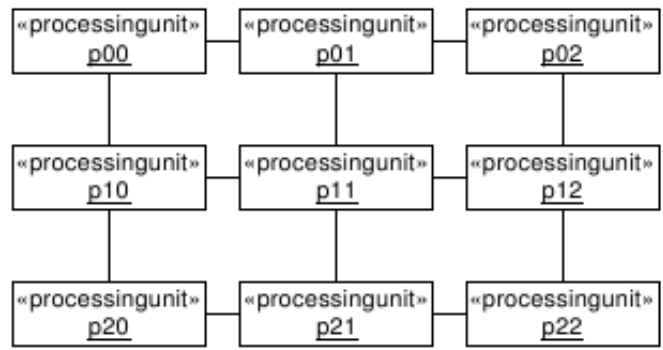

Figure 3: Collaboration Diagram for a Mesh Topology (3x3).

When the developer selects a paradigm to model the sequential algorithm, they should represent the new algorithm. With the objective to describe the parallel solution, developers can describe the logical sequence of processes and events within the parallel algorithm. This visual representation allows to developers to calculate the parallel algorithm complexity, to estimate the cost of performance requirements. By means of the UML diagrams the programmer can optimize the parallel program to find the best parallel design. In the same way that sequential algorithms was described, activity diagrams help to describe the execution flow of the processes and the communications between them (Fig. 6).

In order to represent the parallel region and the sequential region, the diagram represents the different process unit in a different swimlane (Fig. 6). A swimlane is a visual region in an activity diagram that indicates the element that has responsibility for action states within the region. Each swinlane describe the execution work flow from each process unit. The global activity diagram represents the communications between all process units on different times of the execution (Fig. 4). In this representation each swimlane describe a process for the functional decomposition and each swimlane can run in one or several process depending to the process topology. The UML activity diagram to model parallel processes represents the most important activities flows inside the parallel algorithm. The stereotype <<action+>> identifies an atomic action or a block of actions within a sub-functionality of the program. A sub-activity in the diagram describes a sub-function in the main program, it is represent by

<<subactivity+>> stereotype and it related almost always with a function in the algorithm. A control-flow represented by a line transition indicates the order of action states and the object-flow represented by a broken white line describes that an action state inputs or outputs an object. When the algorithm execute a collective communication, the diagram represent a stereotype <<collective+>> connected by a undirected line to a block of send-receive communication. The stereotype <<bsend+>>, <<nbsend+>> identified the type of communication between processes; if the stereotype is preceded by an n is a non-blocking communication. We strongly recommended the usage of notes to describe the actions of the program is a good practice. In (Fig. 4) it was made a remark to the communication actions with the MPI [17] sentence that was used in the algorithm. Moreover, it was considered that this diagram can be described with a simple explanation of the central executions-flow in order to help the UML activity diagram.

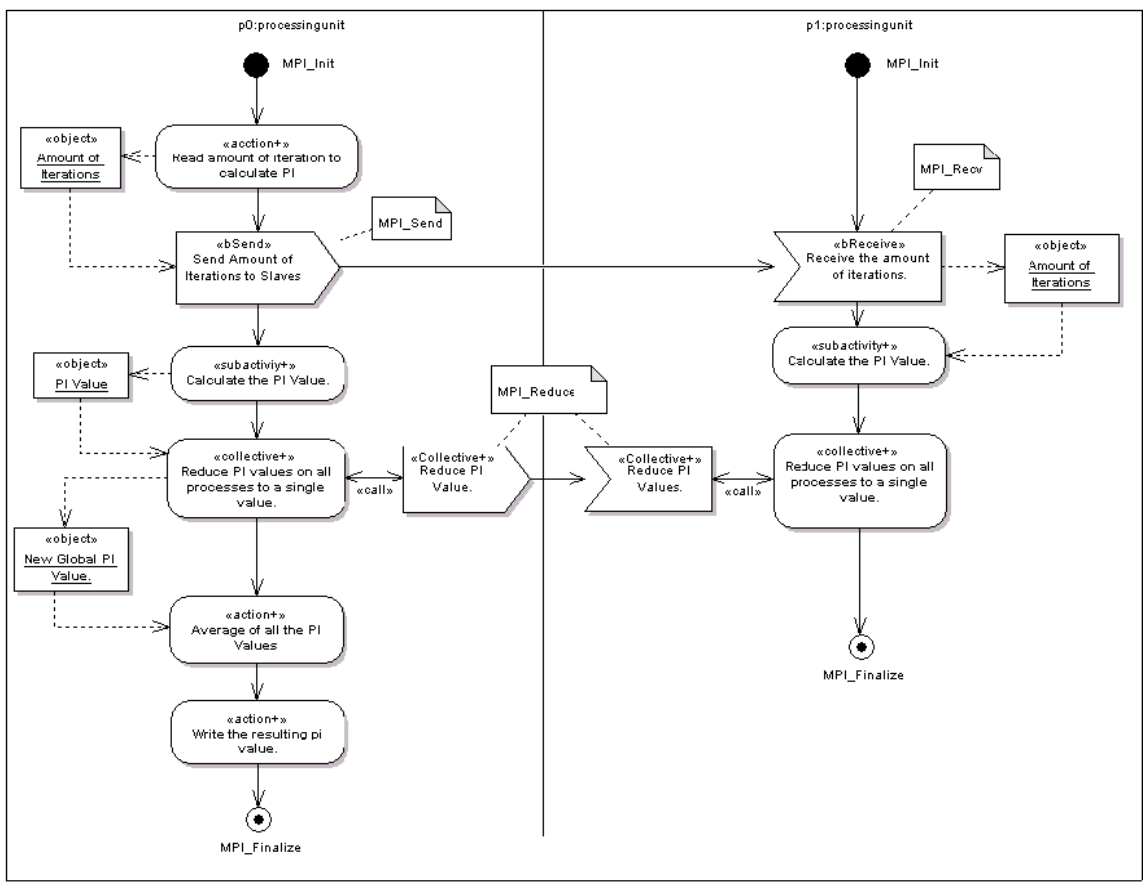

Figure 4: Activity Diagram of computes PI by the Monte Carlo method [18], in a Master-Slave process topology.

Modeling time is an essential element of any real time and/or distributed system. Dealing with real time and distributed systems is hard. Good models reveal the necessary and sufficient properties of a system's time characteristics. A sequence diagram shows elements as they interact over time, showing an interaction or interaction instance. Sequence diagrams are organized along two axes: the horizontal axis shows the elements that are involved in the interaction, and the vertical axis represents time proceeding down the page. It is used to depict work flow; message passing and how elements in general cooperate over time achieve a result [4]. The boxes across the top of the diagram represent the objects, classes. The dashed lines hanging from the boxes are called object lifelines, representing the life span of the object during the scenario being modeled. The long, thin boxes on the lifelines are activation boxes, also called method-invocation boxes, which indicate processing is being performed by the target object/class to fulfill a message. The X at the bottom of an activation box is a UML convention to indicate an object that has been removed from memory (Fig. 5). Process boxes, are represented by stereotypes <<controller>>, indicating that the boxes represent a controller class. These processes can be represented at the beginning of the diagram and before all the others classes or objects. Although in most of the cases of the parallel application only one user interacts with the application, it was represent a simple stereotype actor (<<actor>>) to describe the interaction between the main program and the programmer (Fig. 5). The Main Program Controller represent all the processes in the algorithm, it represent all the functions that execute simultaneously in all the processes including the collective communications (Fig. 5). Common practice on UML diagrams is to indicate creation and destruction messages with the stereotypes of <<create>> and <<destroy>>, respectively. We labels with two possible stereotypes for messages that described an interaction between controller process classes: <<synchronous>> and <<asynchronous>>.

Sequence diagram can be used to predict the performance of the algorithm before the execution of the program on architecture [19, 20]. If the programmer adds to the sequence diagram the time of the instructions and messages, the real time can be calculate. The sequence diagram allows to programmers to change the design of the algorithm depending of the performance requirements.

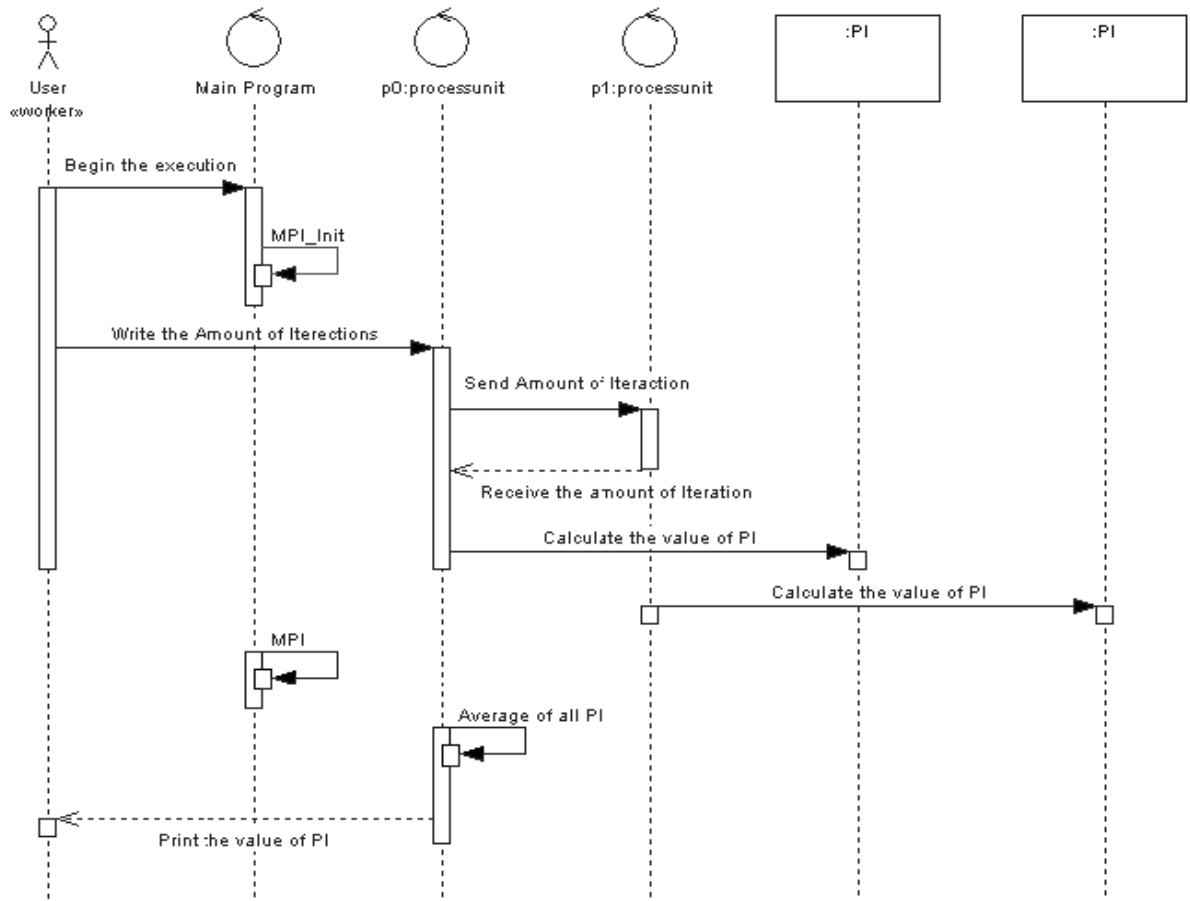

Figura 5: Sequence Diagram of computes PI by the Monte Carlo method [18], in a Master-Slave process topology.

**Conclusions**

Systems being developed now are more complex than ever, and old software development methods simply do not efficiently scale up to the size of current system. The Unified Modeling Language (UML) is a standard widely-adopted graphical language that describes the artifacts of software system with a focus on conceptual and physical representations. UML is effective for modeling large, complex software systems. It is simple to learn for most developers, but provides advanced features for expert analysts, designers and architects.

In this paper, we have described an UML approach to model parallel and distributed applications. We present some phases of the development of the software to model parallel programs. We focused in the domain model to introduced the developers in the sequential problem and solution, and describe the principals requirements for a parallel solution. We presented several programming paradigms are commonly used to develop parallel programs on distributed clusters and the cloud. UML provides a common language to developer teams and a set of modeling concepts, notations and mechanisms to substantially alleviate the understanding, documentation, and visualization of the parallel applications. This manuscript provides a guide and phases to design and modeling a parallel and distributed application from the problem to the final parallel solution.

**References**

[1] Schaller RR. Moore's law: past, present and future. Spectrum, IEEE. 1997;34:52-9.
[2] Wilkinson B, Allen CM. Parallel programming: Prentice hall New Jersey; 1999.
[3] Foster I. Designing and building parallel programs: Addison-Wesley Reading; 1995.
[4] Goswami D, Singh A, Preiss B. Building Parallel Applications Using Design Patterns. In: Erdogmus H, Tanir O, editors. Advances in Software Engineering: Springer New York; 2002. p. 243-65.
[5] Bennett S, McRobb S, Farmer R. Object-oriented systems analysis and design using UML: McGraw-Hill Berkshire,, UK; 2006.
[6] Pllana S, Fahringer T. Performance prophet: A performance modeling and prediction tool for parallel and distributed programs. Parallel Processing, 2005 ICPP 2005 Workshops International Conference Workshops on: IEEE; 2005. p. 509-16.
[7] Pllana S, Fahringer T. UML based modeling of performance oriented parallel and distributed applications. Simulation Conference, 2002 Proceedings of the Winter: IEEE; 2002. p. 497-505.
[8] Kruchten P. The rational unified process: an introduction: Addison-Wesley Professional; 2004.
[9] Martin RC. Agile software development: principles, patterns, and practices: Prentice Hall PTR; 2003.
[10] Cockburn A. Crystal clear: a human-powered methodology for small teams: Pearson Education; 2004.
[11] Jackson M. Software requirements & specifications: ACM Press New York; 1995.
[12] Yu ES. Towards modelling and reasoning support for early-phase requirements engineering. Requirements Engineering, 1997, Proceedings of the Third IEEE International Symposium on: IEEE; 1997. p. 226-35.
[13] Hendrix D, Cross J, Maghsoodloo S. The effectiveness of control structure diagrams in source code comprehension activities. Software Engineering, IEEE Transactions on. 2002;28:463-77.
[14] Pllana S, Fahringer T. On customizing the UML for modeling performance-oriented applications. ≪ UML≫ 2002—The Unified Modeling Language: Springer; 2002. p. 259-74.
[15] Leopold C. Parallel and Distributed Computing: A survey of models, paradigms, and approaches: Wiley; 2001.
[16] Gomaa H. Designing concurrent, distributed, and real-time applications with UML. Proceedings of the 28th international conference on Software engineering: ACM; 2006. p. 1059-60.
[17] Gropp WD, Lusk EL, Skjellum A. Using MPI: portable parallel programming with the message-passing interface: the MIT Press; 1999.
[18] Perez-Riverol Y, Vera R, Mazola Y, Musacchio A. A parallel systematic-Monte Carlo algorithm for exploring conformational space. Curr Top Med Chem. 2012;12:1790-6.
[19] Briand LC, Labiche Y, Leduc J. Toward the reverse engineering of UML sequence diagrams for distributed Java software. Software Engineering, IEEE Transactions on. 2006;32:642-63.
[20] Li X, Lilius J. Timing analysis of UML sequence diagrams: Springer; 1999.